\documentclass[reprint,longbibliography,aip]{revtex4-2}

\usepackage{graphicx} 
\usepackage{amsmath}
\usepackage{ amssymb }

\usepackage{color}

\usepackage{float} 
\usepackage{wrapfig} 

\usepackage[makeroom]{cancel} 
\usepackage{comment}

\usepackage{lipsum} 

\usepackage{outlines}

\graphicspath{{./figs/}} 

\newcommand{\beq}{\begin{equation}}
\newcommand{\eeq}{\end{equation}}

\newcommand{\bvec}{\begin{pmatrix}}
\newcommand{\evec}{\end{pmatrix}}
\newcommand{\lp}{\left(}
\newcommand{\rp}{\right)}

\newcommand{\llangle}{\left \langle}
\newcommand{\rrangle}{\right \rangle}

\AtBeginDocument{%
	\newwrite\bibnotes
	\def\bibnotesext{Notes.bib}
	\immediate\openout\bibnotes=\jobname\bibnotesext
	\immediate\write\bibnotes{@CONTROL{REVTEX41Control}}
	\immediate\write\bibnotes{@CONTROL{%
			apsrev42Control,author="08",editor="1",pages="1",title="0",year="1"}}
	\if@filesw
	\immediate\write\@auxout{\string\citation{apsrev42Control}}%
	\fi
}%

\begin{document}



\title{Lowering the reactor breakeven requirements for proton-Boron 11 fusion}

\author{Ian E. Ochs}
\affiliation{Department of Astrophysical Sciences, Princeton University, Princeton, New Jersey 08540, USA}
\author{Nathaniel J. Fisch}
\affiliation{Department of Astrophysical Sciences, Princeton University, Princeton, New Jersey 08540, USA}

\date{\today}

\begin{abstract}

Recently, it has been shown that altering the natural collisional power flow of the proton-Boron 11 (pB11) fusion reaction can significantly reduce the Lawson product of ion density and confinement time required to achieve ignition.
However, these products are still onerous---on the order of $7\times 10^{15}$ cm$^{-3}$s under the most optimistic scenarios.
Fortunately, a breakeven fusion power plant does not require an igniting plasma, but rather a reactor that produces more electrical power than it consumes.
Here, we extend the existing 0D power balance analysis to check the conditions on power plant breakeven.
We find that even for the base thermonuclear reaction, modern high-efficiency thermal engines should reduce the Lawson product to $1.2\times 10^{15}$ cm$^{-3}$s.
We then explore the impact of several potential improvements, including fast proton heating, alpha power capture, direct conversion, and efficient heating.
We find that such improvements could reduce the required Lawson product by a further order of magnitude, bringing aneutronic fusion to within target ITER design parameters.

\end{abstract}

\maketitle


\section{Introduction} 

While the deuterium-tritium (DT) reaction has historically dominated fusion research efforts because of its high cross section at relatively low temperature, the fact that it relies on scarce, radioactive tritium and produces 14 MeV fast neutrons introduce large technological and market fit risks.
This has recently led to a resurgence of interest in abundant, aneutronic fuels, such as proton-Boron 11 (pB11).

However, pB11 is a much harder reaction than DT, occurring with smaller cross sections at much greater temperatures.
For a long time, it was thought that a self-sustaining reaction would be impossible to produce in a pB11 plasma, since the bremsstrahlung power would always exceed the fusion power.\cite{Rider1995,Rider1995a}
Fortunately, this assertion was based in part on overly low cross sections,\cite{Nevins2000CrossSection} and newer cross section data\cite{Sikora2016CrossSection} opened up a narrow window of feasibility for thermonuclear pB11 ignition.\cite{Putvinski2019}
However, this ignition window required extremely long confinement times, on the order of 450 seconds at ion densities of $10^{14}$ cm$^{-3}$.\cite{Ochs2022ImprovingFeasibility}

The thermonuclear power balance analysis assumes that power is transferred collisionally from the fusion-born alpha particles into the various species in the plasma.
However, this need not be the case---one can alter the natural flow of energy in the plasma by either co-localizing certain species, or enhancing the energy transfer rate via waves.\cite{Fisch1992,Valeo1994,Herrmann1997}
Recent work\cite{Ochs2022ImprovingFeasibility,kolmes2022waveSupported} has shown that altering the power balance by transferring more alpha energy directly into fast protons could substantially increase the fusion power and reduce the bremsstrahlung power, reducing the confinement time needed for ignition by a factor of more than 6, to 70 seconds at ion densities of $10^{14}$ cm$^{-3}$.

A 70 second confinement time is still quite prohibitive.
However, this only represents the confinement time needed for a self-sustaining fusion reaction, without external heating. 
Such a reaction is not strictly necessary for the achievement of a breakeven fusion power plant.
Instead, all that is required is that the electrical power out exceed the electrical power in.
Indeed, the pB11 literature is characterized by interest in nonthermal fusion schemes with external heating. \cite{Rostoker1997BeamPB11,LampeMannheimer1998CommentsPB11,Volosov2006ACT,Volosov2011Problems,Labaune2013LaserPB11,Ruhl2022LaserPB11,Eliezer2016Avalanche,Magee2019DirectObservation,eliezer2020NovelFusion,Eliezer2020BeamPB11,Istokskaia2023MultiMeVAlpha,Wei2023ProtonBoronFusion,Magee2023FirstMeasurements}

In this paper, we extend the existing 0D power balance analysis\cite{Ochs2022ImprovingFeasibility} to account for the possibility of external heating, and optimize for the power-pant-relevant metric $Q_\text{eng}=(P_\text{out} - P_\text{in})/P_\text{in}$.
This extension allows us to test the impacts of various possible approaches to reducing the energy confinement time required for breakeven, including targeting fast particles for heating, capture power from alpha particles into fast protons, more efficiently delivering heating power to the reactor, and using direct conversion to efficiently recapture power from lost particles.

\section{Reactor Power Flow and Q}\label{sec:performance}

\begin{figure}[t]
	\centering
	\includegraphics[width=0.8\linewidth]{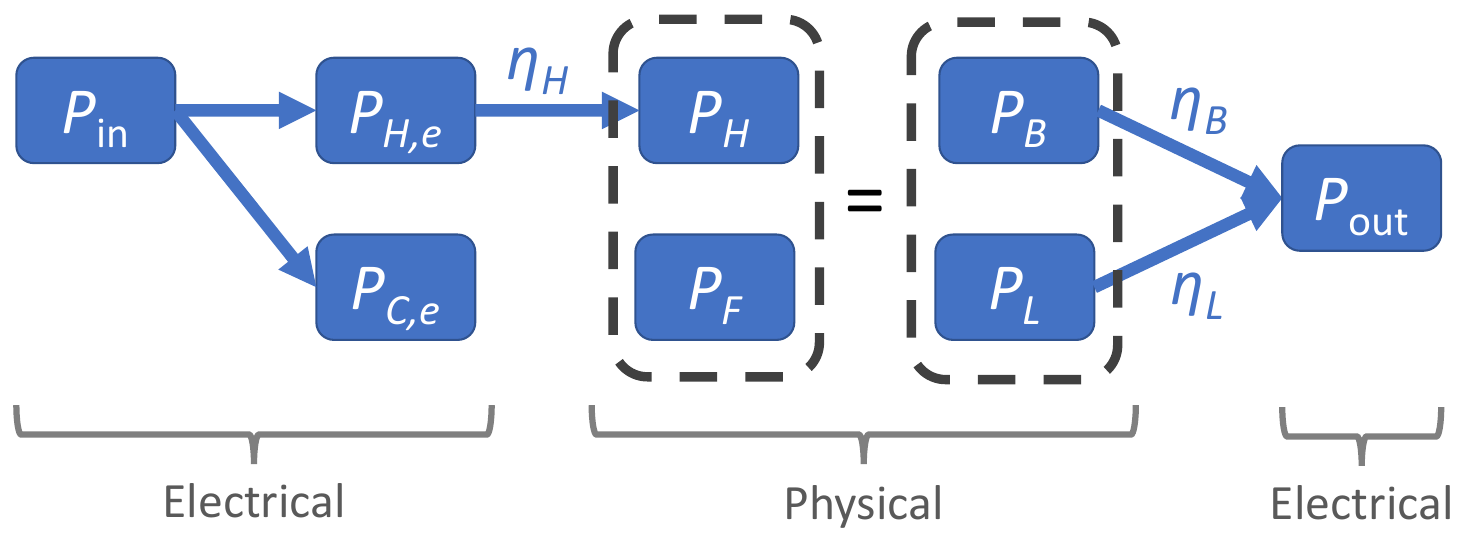}
	\caption{Reactor power flow, as described in Sec.~\ref{sec:performance}. Reactor feasibility requires $P_\text{out} > P_\text{in}$, which forms the basis for the parameter $Q_\text{eng}$.
	Different values of heating efficiency $\eta_H$, bremsstrahlung radiation $\eta_B$, and thermal conduction efficiency $\eta_L$ generally result in different values of $Q_\text{eng}$.}
	\label{fig:PlantPowerBalance}
\end{figure}

In this section, we quickly review the power flow model (Fig.~\ref{fig:PlantPowerBalance}) from Ref.~\onlinecite{Ochs2022ImprovingFeasibility}, which is related to that from Ref.~\onlinecite{WurzelHsu2022Progress}, and which forms the basis for the analysis in this paper.
In the reactor power flow, electrical power $P_\text{in}$ consists both of power used to heat ($P_{H,e}$) and confine ($P_{C,e}$) the plasma.
With a conversion efficiency $\eta_H$, this electrical heating power is delivered to the plasma as heating power $P_H = \eta_H P_{H,e} $.
At the same time, the plasma produces some amount of fusion power $P_F$.
Power exits the plasma through two possible mechanisms: bremsstrahlung radiation $P_B$, or thermal conduction loss $P_L$.
In steady state:
\begin{align}
	P_H + P_F = P_B + P_L. \label{eq:PBalanceReactionAbstract}
\end{align}
The relationship between these terms is determined by the internal power balance equations.
In keeping with past work, we will often write the thermal conduction losses in terms of the confined kinetic energy density $U_K$ and the energy confinement time $\tau_E$:
\begin{align}
	\tau_E \equiv U_K/P_L. \label{eq:tauEDef}
\end{align}
At the exhaust, the power that exits the plasma is converted back to into electrical power, with in general different efficiencies $\eta_B$ and $\eta_L$ for bremsstrahlung and thermal conduction losses respectively.
This results in a final output electrical power $P_\text{out} = \eta_B P_B + \eta_L P_L$.
Economical fusion energy requires that $P_\text{out}$ exceed $P_\text{in}$.
In the limit of the confinement system power going to 0, this condition requires $Q_\text{eng} > 0$, where
\begin{align}
	Q_\text{eng} &\equiv \frac{P_\text{out} - P_\text{in}}{P_\text{in}} = \frac{\eta_H\left(\eta_B P_B + \eta_L P_L\right)}{P_H} - 1. \label{eq:QengStar}
\end{align}
In this paper, we will determine the maximum value of $Q_\text{eng}$, which we denote $Q_\text{eng}^*$, which is achievable for given constraints on the various efficiencies $\eta$, as well as on internal power balance parameters, discussed in Section~\ref{sec:powerBalanceModel}.

\section{Internal Power Balance}\label{sec:powerBalanceModel}

In this section, we briefly review the internal power balance model from Ref.~\onlinecite{Ochs2022ImprovingFeasibility}, which the reader should refer to for more details.
The power balance equations for a fast protons $f$, thermal protons $p$, thermal boron $b$, and thermal electrons $e$ is given by:
\begin{align}
	\frac{dU_f}{dt} &= - K_{f p} E_f  - K_{f b} E_f - K_{f e} E_f \notag\\ 
	&\hspace{0.25in}- K_{F,f} E_f + \alpha_f P_\alpha + \chi P_H - \gamma_f P_L \label{eq:dEfdt}\\
	\frac{dU_p}{dt} &= K_{f p} E_f + K_{pb} (T_b - T_p) + K_{p e} (T_e - T_p) \notag\\ 
	&\hspace{0.25in} - \frac{3}{2} K_{F,p} T_p + \alpha_p P_\alpha + (1-\chi) P_H -\gamma_p P_L\\
	\frac{dU_b}{dt} &= K_{f b} E_f + K_{pb} (T_p - T_b) + K_{b e} (T_e - T_b)  \notag\\ 
	&\hspace{0.25in}-\frac{3}{2} (K_{F,f}+K_{F,p}) T_b + \alpha_b P_\alpha - \gamma_b P_L\\
	\frac{dU_e}{dt} &= K_{f e} E_f + K_{pe} (T_p - T_e) + K_{b e} (T_b - T_e) \notag\\ 
		&\hspace{0.25in}- P_B + \alpha_e P_\alpha. \label{eq:dTedt}
\end{align}
Here, we recognize the heating power $P_H$, thermal conduction loss power $P_L$, and bremsstrahlung power $P_B$.

\begin{figure}[t]
	\centering
	\includegraphics[width=\linewidth]{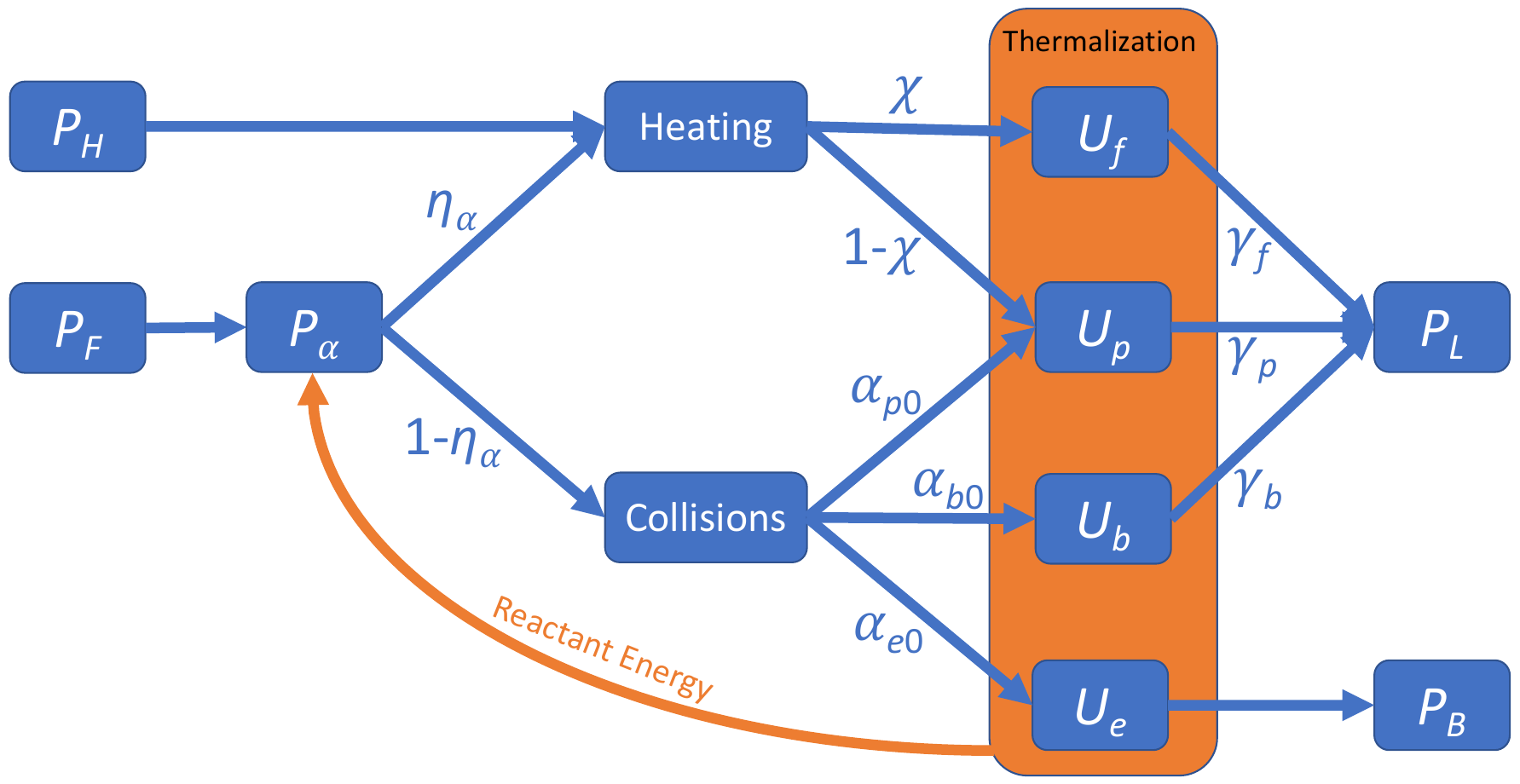}
	\caption{Internal power balance, as described in Sec.~\ref{sec:powerBalanceModel}.
	Different values of alpha power capture efficiency $\eta_\alpha$ and fast particle heating $\chi$ result in different relationships between $P_H$, $P_F$, $P_L$, and $P_B$.}
	\label{fig:InternalPowerBalance}
\end{figure}

The various variables are described in detail in Ref.~\onlinecite{Ochs2022ImprovingFeasibility}.
In quick summary,
$K_{ss'}$ for $s,s' \in \{f,p,b,e\}$ represents the energy transfer rate between species $s$ and $s'$. 
$K_{F,f}$ and $K_{F,p}$ represent the fusion rate from fast and thermal protons respectively.
$\gamma_s$ represents the proportion of thermal conduction losses from species $s$.
The power $P_\alpha$ represents the power flowing through the alphas from both the reactant kinetic energy and the fusion energy ($\mathcal{E}_F = 8.7$ MeV):
\begin{align}
	P_\alpha &\equiv \mathcal{E}_F (K_{F,f} + K_{F,p}) + K_{F,f} \lp E_f + \frac{3}{2} T_b \rp\notag\\
	& \hspace{1.13in} + \frac{3}{2} K_{F,p} (T_p + T_b),
\end{align}

The terms that alter the internal power balance reside in the coefficients $\chi$ and $\alpha_s$.
First, we assume that some fraction of the heating power $\chi$ ends up in the fast protons $f$, with $(1-\chi)$ going to the thermal protons $p$.
Then, we assume that some determined fraction $\eta_\alpha$ of the alpha power can be redirected as heating power into the protons, again apportioned between fast and thermal protons according to $\chi$.
The remaining alpha particle energy is partitioned between species according to the collisional slowing down on those species, which is captured in the parameter:
\begin{align}
	\alpha_{s0} \equiv \llangle \frac{K_{\alpha s}}{K_{\alpha p} + K_{\alpha b} + K_{\alpha e}} \rrangle, \quad s \in \{p,b,e\}.
\end{align}
Here, the average is performed appropriately over the hot alpha particle distribution.\cite{Ochs2022ImprovingFeasibility}
Thus, the total fraction of alpha particle power going to each species, including both altered power flow and collisional effects, is given by:
\begin{align}
	\alpha_f &= \eta_\alpha \chi \label{eq:alphaF}\\
	\alpha_p &= (1-\eta_\alpha) \alpha_{f0} + \eta_\alpha(1-\chi)\\
	\alpha_b &= (1-\eta_\alpha) \alpha_{b0}\\
	\alpha_e &= (1-\eta_\alpha) \alpha_{e0}. \label{eq:alphaE}
\end{align}

There are a couple changes to this model relative to Ref.~\onlinecite{Ochs2022ImprovingFeasibility}.
First, the bremsstrahlung power is given by:\cite{Heitler2012Radiation,Ochs2022ImprovingFeasibility,Putvinski2019,Munirov2023SuppressionBremsstrahlung}
\begin{align}
	P_B &\approx 7.56 \times 10^{-11} n_e^2 x^{1/2} \bigl[ Z_\text{eff} \lp 1 + 1.78x^{1.34} \rp \notag\\
	& \hspace{0.2in} + 2.12 x \lp 1 + 1.1 x + x^2 - 1.25 x^{2.5} \rp \bigr] \text{ eV cm$^{3}$/s,} \label{eq:PbrFormula}
\end{align}
where $x = T_e / E_\text{rest}$, $Z_\text{eff} = \sum_i n_i Z_i^2 / \sum_i n_i Z_i$, and $E_\text{rest} = 5.11\times 10^5$ eV is the electron rest energy.
Note that Ref.~\onlinecite{Ochs2022ImprovingFeasibility} missed the $x^2$ term in the second bracket.
As a result, the kinetic enhancement factor for the fusion rate also had to be slightly changed to agree with Ref.~\onlinecite{Putvinski2019}, from a maximum $\phi_k(0 \text{ keV}) = 1.16$  to $\phi_k(0 \text{ keV}) = 1.178$.

Second, in this paper, we assume that thermal losses in each ion species occur proportionally to that species' internal energy, including for the fast proton species.
Thus, 
\begin{align}
	\gamma_i \equiv \frac{U_i}{\sum_{j} U_j}, \quad i,j\in \{f,p,b\},
\end{align}
where $U_i = \frac{3}{2} n_i T_i$ for $i \in \{p,b\}$, while $U_f = n_f E_f$.

The overall power flow represented by Eqs.~(\ref{eq:dEfdt}-\ref{eq:dTedt}) and (\ref{eq:alphaF}-\ref{eq:alphaE}), incorporating both collisions and alpha channeling, is schematically represented in Fig.~\ref{fig:InternalPowerBalance}.

\section{Reactor Performance Optimization}\label{sec:optimization}

To measure reactor performance, our goal is to find the optimal $Q_\text{eng}^*$ for given $\eta_H$, $\eta_L$, $\eta_B$, $\eta_\alpha$, $\chi$, and $\tau_E$.
We perform a constrained optimization  over the four variables $n_p$, $n_b$, $E_f$, and $P_H$, with the remaining variables ($n_e$, $T_p$, $T_b$, and $T_e$) determined by the power balance Eqs.~(\ref{eq:dEfdt}-\ref{eq:dTedt}) and quasineutrality.
The optimization is constrained so that $n_i \equiv n_f+n_p+n_b = 10^{14}$ cm$^{-3}$.
The optimization is performed using the trust region algorithm \cite{Conn2000TrustRegion} for constrained optimization (trust-constr) implemented as an option in to scipy.minimize.
Note that in contrast to Ref.~\onlinecite{Ochs2022ImprovingFeasibility}, $P_L$ is now a fixed function of the other variables based on the given confinement time.


First, we note that a separate optimization is not required for each value of $\eta_H$ and $\eta_B$, since (a) $\eta_H$ and $\eta_B$ do not enter the internal power balance, and (b) from Eq.~(\ref{eq:QengStar}), it is clear that $Q_{eng}^*$ is maximized for a given value of $\eta_H$ and $\eta_B$ when $\left(P_B + (\eta_L/\eta_B) P_L\right)/P_H$ is maximized.
Taking $\bar{\eta}_H$ and $\bar{\eta}_B$ as the values of $\eta_H$ and $\eta_B$ for the simulated case, the value of $Q_\text{eng}^*$ for other values is:
\begin{align}
	Q_\text{eng}^*\bigl|_{\eta_H, \eta_B} = \frac{\eta_H \eta_B}{\bar{\eta}_H \bar{\eta}_B}\lp Q_\text{eng}^*\bigl|_{\bar{\eta}_H,\bar{\eta}_B} + 1\rp - 1.
\end{align}
Thus, we only need to run separate optimization simulations for different sets of parameters $\eta_\alpha$, $\chi$, $\tau_E$, and the ratio $\eta_L/\eta_B$.

\begin{figure}[t]
	\centering
	\includegraphics[width=\linewidth]{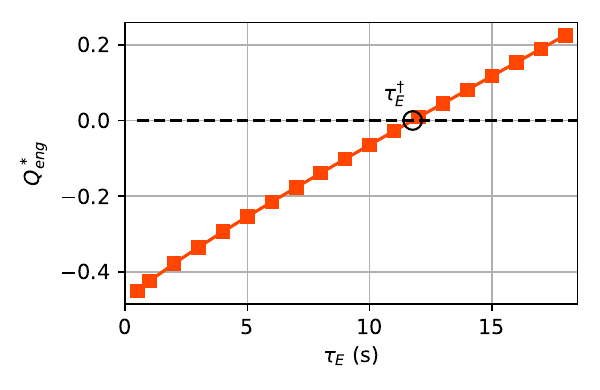}
	\caption{Plot of $Q_\text{eng}^*$ vs. $\tau_E$ for base case of thermonuclear p-B11 fusion at an ion density of $10^{14}$ cm$^{-3}$.
		The reactor breakeven confinement time $\tau_E^\dagger$ at which electrical power out exceeds electrical power in has a value of 11.8 seconds for this reaction.}
	\label{fig:QVsTauEThermonuclear}
\end{figure}

\subsection{Thermonuclear base case and breakeven confinement time}

For the thermonuclear base case, we assume bremsstrahlung and thermal conduction losses are converted into electricity with 64\% efficiency, in line with state-of-the-art gas turbine technology.
We also assume $\chi = 0$ and $\eta_\alpha=0$, and a base heating efficiency of $\eta_H = 0.8$ (slightly more optimistic than the $\eta_H = 0.7$ used in Ref.~\onlinecite{WurzelHsu2022Progress}).
We can then scan $Q_\text{eng}^*$ as a function of $\tau_E$, to get a mapping of reactor performance as a function of confinement time (at $n_i = 10^{14}$ cm$^{-3}$).
This plot is shown in Fig.~\ref{fig:QVsTauEThermonuclear}.

The scan of $Q_\text{eng}^*$ vs. $\tau_E$ allows us to define an important new quantity: the reactor breakeven confinement time $\tau_E^\dagger$, at which the electrical power out exceeds the electrical heating power in.
This is given by the point at which $Q_\text{eng}^*(\tau_E^\dagger) = 0$, as represented by the circle in Fig.~\ref{fig:QVsTauEThermonuclear}.
For the thermonuclear base case, this breakeven confinement time is 11.8 seconds.
Note that $\tau_E^\dagger$, the reactor breakeven time, is a different (and generally much smaller) quantity than $\tau_E^*$, the ignition breakeven time.\cite{Ochs2022ImprovingFeasibility,kolmes2022waveSupported}

\subsection{Improving on the thermonuclear baseline}\label{sec:IndividualScenarios}

With the base case established, we can begin to consider the impact of additional strategies.
First, we can consider fast proton heating (FPH), i.e. directing heating power into the beam of high-energy protons.
For such a scenario, we take $\chi = 0.8$, instead of the base case of $\chi = 0$.

Second, we can consider the effect of altering the power flow from alpha particles into protons, which we term alpha power capture (APC).
In such a case, we take $\eta_\alpha = 0.8$, instead of $\eta_\alpha = 0$.
Note that the definitions of $\eta_\alpha$ and $\chi$ imply that in scenarios with both FPH and APC, the power directly channeled into fast protons from alphas is $\eta_\alpha \chi P_\alpha = 0.64 P_\alpha$ (see Eq.~(\ref{eq:alphaF})).

\begin{table}[t]
	\centering
	\begin{tabular}{|c|c|c|c|}
		\hline
		\, Symbol \, & Meaning & Definition& Base \\
		\hline
		FPH &  Fast Proton Heating & $\chi = 0.8$& \,$\chi = 0$\,\\
		APC & \, Alpha Power Capture \, & $\eta_\alpha = 0.8$ & \,$\eta_\alpha = 0$\,\\
		DC & Direct Conversion & \, $\eta_L = \tfrac{4}{3} \eta_B$ \, & \, $\eta_L = \eta_B$ \,\\
		EH & Efficient Heating & $\eta_H = 0.95$& $\eta_H = 0.8$\\
		\hline
	\end{tabular}
	\caption{The meaning of the various strategies for increasing reactor performance in terms of the power balance variables, compared to the variables in the base thermonuclear case.}
	\label{tab:StrategySymbols}
\end{table}

Third, we examine the effect that direct conversion (DC) might play on the power balance.
In direct conversion, some fraction of the charged particle power flowing out of the plasma is captured as electrical potential energy, before the remainder is converted to thermal energy.
Schemes for this include the Venetian blind \cite{Moir1973VenetianblindDirect} or concentric end electrode traps. \cite{Volosov2006ACT,Volosov2011Problems}
The efficiency of the direct conversion need not be particularly high to significantly improve the overall electrical conversion efficiency $\eta_L$, since the total conversion efficiency is:
\begin{align}
	\eta_L = \eta_\text{DC} + (1-\eta_\text{DC}) \eta_\text{Th},
\end{align}
where $\eta_\text{DC}$ and $\eta_\text{Th}$ are the electrical conversion energies from direct conversion and thermal cycles respectively.
So, for instance, if $\eta_\text{DC} = 0.6$ and $\eta_\text{Th} = \eta_B = 0.64$, then $\eta_L = 0.86 \approx \tfrac{4}{3} \eta_B$.
Thus, we take our direct conversion scenario to be $\eta_L = \tfrac{4}{3} \eta_B$, rather than the base case.

Finally, we can also look at the impact of more efficient heating (EH). 
To do this, we simply take $\eta_H = 0.95$ rather than the base case of $\eta_H = 0.8$.
These various strategies, along with their meanings, mathematical definitions, and comparison to the base thermonuclear scenario, are collected in Table~\ref{tab:StrategySymbols}.

\begin{figure}[t]
	\centering
	\includegraphics[width=\linewidth]{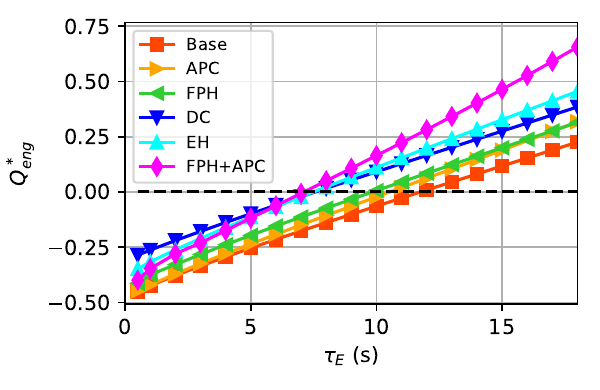}
	\caption{Reactor performance $Q_\text{eng}^*$ as a function of confinement time $\tau_E$ for the strategies presented in Table~\ref{tab:StrategySymbols}, as well as for a combination of fast proton heating (FPH) and alpha power capture (APC).
		While FPH and APC each produce a modest performance increase individually, the combination strategy is greater than the sum of the parts.}
	\label{fig:QVsTauESingle}
\end{figure}

In Fig.~\ref{fig:QVsTauESingle}, we see the impact of each individual strategy on the reactor performance $Q_\text{eng}^*(\tau_E)$.
All strategies improve the reactor performance, but DC and EH, at the levels given in Table~\ref{tab:StrategySymbols}, have the biggest individual impact.
However, it is also notable that the performance of a combination of FPH and APC is much greater than the sum of the parts, with each individually only giving an increase of $\Delta Q_\text{eng}^*\approx 0.1$, while the combination gives and increase of up to $\Delta Q_\text{eng}^*\approx 0.5$.
This is reminiscent of the results of the ignition calculation of Refs.~\onlinecite{Ochs2022ImprovingFeasibility,kolmes2022waveSupported}, where transferring alpha particle energy into tail protons resulting in a much larger reduction in the confinement time $\tau_E^*$ required for \emph{ignition} than transferring alpha particle energy into thermal protons.

\begin{figure}[t]
	\centering
	\includegraphics[width=\linewidth]{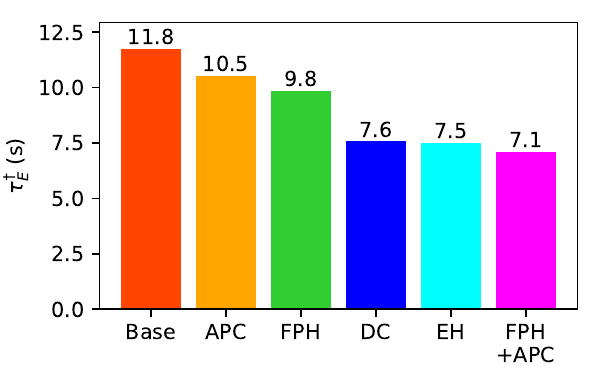}
	\caption{Reactor breakeven confinement time $\tau_E^\dagger$ (see Fig.~\ref{fig:QVsTauEThermonuclear}) for different scenarios from Table~\ref{tab:StrategySymbols}.
		DC and EH have the greatest individual impacts, but the combination of FPH and APC is also more powerful than either strategy individually.}
	\label{fig:TauBreakevenSingle}
\end{figure}

We can also look at the different strategies in terms of their reduction of the reactor breakeven confinement time $\tau_E^\dagger$, which are shown in Fig.~\ref{fig:TauBreakevenSingle}.
These show that, as for $Q^*_\text{eng}$, DC and EH have the biggest individual impacts, with $\Delta \tau_E^\dagger \approx -4.2$ relative to the base case, compared to $\Delta \tau_E^\dagger \approx -1.3$ for APC and $\Delta \tau_E^\dagger \approx -2$ for FPH.
Again, however, the FPH and APC synergize, so that the combination of both strategies produces a reduction $\Delta \tau_E^\dagger \approx -4.6$.

\begin{figure*}[t]
	\centering
	\includegraphics[width=\linewidth]{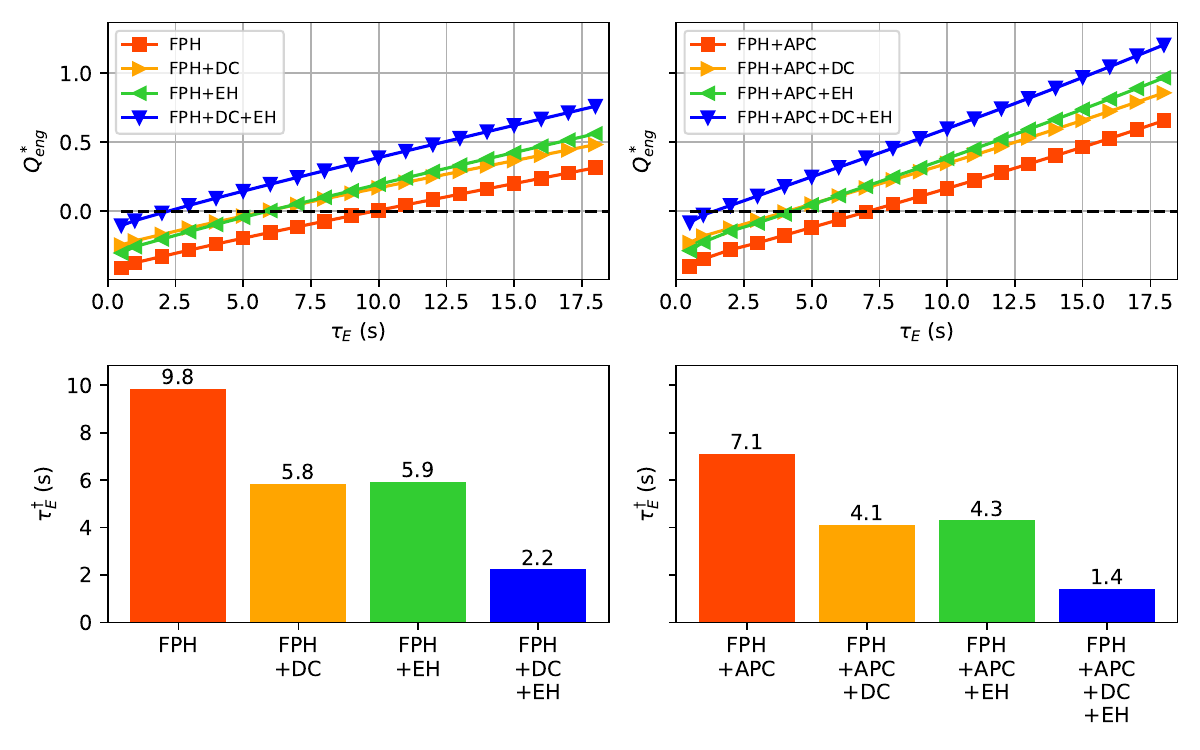}
	\caption{Reactor performance $Q_\text{eng}^*$ (top) and reactor breakeven time $\tau_E^\dagger$ (bottom) for combinations of the strategies in Table~\ref{tab:StrategySymbols}.
		On the left FPH is the base case, and DC and EH are added, while on the right FPH + APC is the base case.
		Generally, FPH + APC leads to approximately a 30\% reduction in $\tau_E^\dagger$ relative to FPH alone.
		DC and EH each provide around a 40\% reduction individually, which stack straightforwardly, producing an 80\% reduction in combination.
		Use of all four strategies reduces the breakeven confinement time by almost an order of magnitude, from 11.8 seconds in the thermonuclear base case to 1.4 seconds.}
	\label{fig:QAndTauEStStCombo}
\end{figure*}

The extreme nonlinearity of the system makes a linear sensitivity analysis rather crude, as underlined by the fact that the combination APC+FPH strategy gives such a different result that the sum of each individual strategy.
Nevertheless, one can translate the simulation results into a linear model to get a very rough idea of the impact of each term on reactor performance, simply by dividing the change in $\tau_E^\dagger$ by the change in the parameter that led to it.
This gives:
\begin{align}
	\tau_E^\dagger \sim 11.8 - 1.5 \eta_\alpha - 2.4 \chi - 20(\eta_L - 0.64) - 28(\eta_H - 0.8). \label{eq:linearSensitivity}
\end{align}
It is important to remember that the large coefficients in front of $\eta_L$ and $\eta_H$ tend to multiply quantities which can only change by $\mathcal{O}(0.1)$, while $\eta_\alpha$ and $\chi$ can change by $\mathcal{O}(1)$.

\subsection{Combination strategies}

Section~\ref{sec:IndividualScenarios} made it clear that using combinations of strategies can produce quite different results from considering the strategies individually.
Thus, in this section, we look at how various combinations of the strategies in Table~\ref{tab:StrategySymbols} can impact the reactor performance.

For combinatoric organization, we consider two basic base scenarios: FPH alone, and FPH + APC.
To each of these base scenarios, we then add DC, EH, and finally DC + EH.
The results are summarized in Fig.~\ref{fig:QAndTauEStStCombo}, with base FPH on the left and FPH + APC on the right, and $Q_\text{eng}^*(\tau_E)$ on the top and $\tau_E^\dagger$ on the bottom. 

Some general trends are clearly visible from the plots.
First, regardless of DC or EH, the combination of FPH + APC leads to about a 30\% reduction in $\tau_E^\dagger$ relative to FPH alone.
Second, DC and EH each individually reduce $\tau_E^\dagger$ by around 40\%, and in combination lower $\tau_E^\dagger$ by around 80\%.
These results suggests that the synergy observed between FPH and APC is somewhat unique in the parameter space, while other effects stack more straightforwardly.
As a net result of using all the strategies, the energy confinement time $\tau_E^\dagger$ can be reduced by almost an order of magnitude relative to the thermonuclear base case, i.e. from 11.8 seconds to 1.4 seconds.

\begin{figure*}
	\centering
	\includegraphics[width=\linewidth]{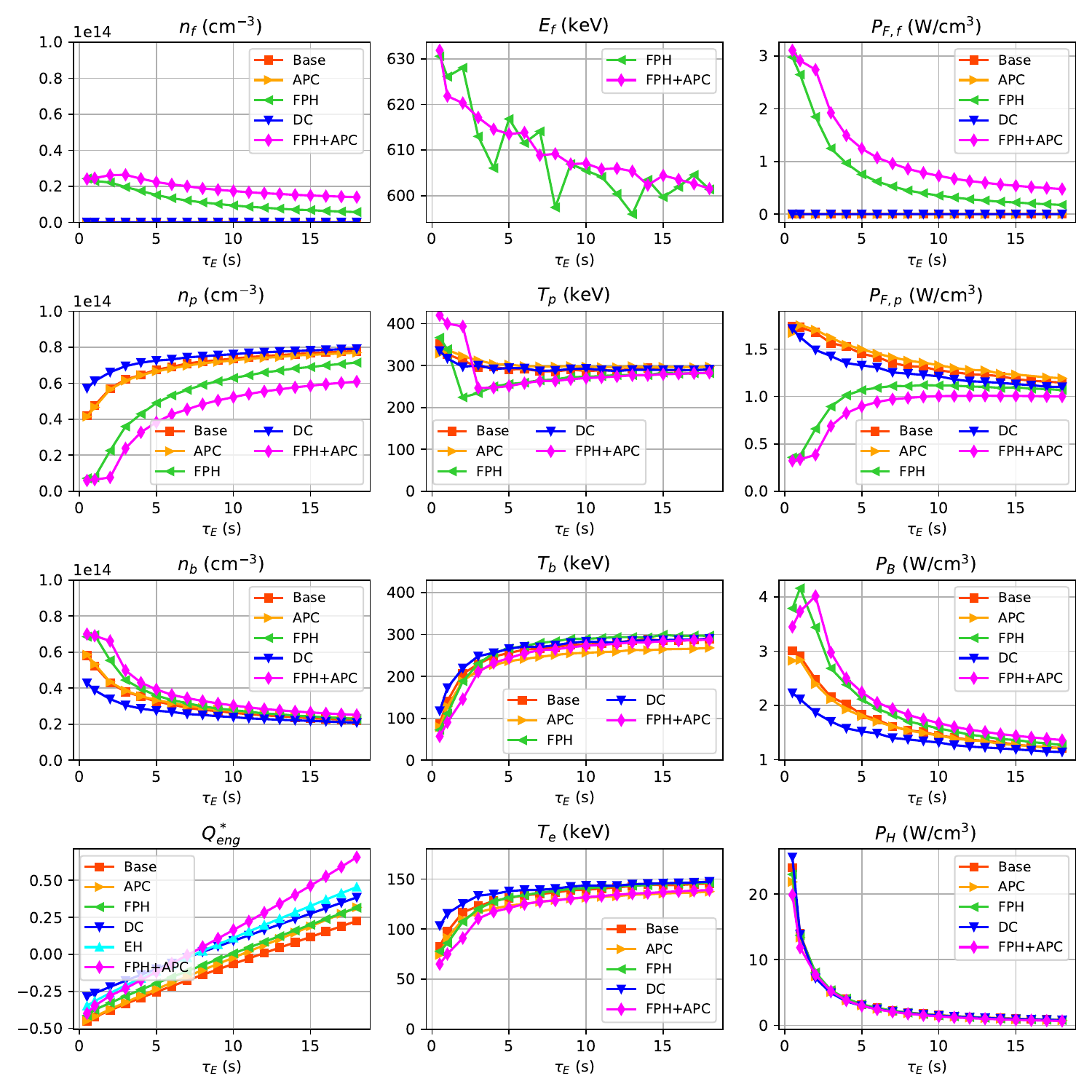}
	\caption{Optimal reactor parameters as a function of $\tau_E$ for the single-scenario runs from Table~\ref{tab:StrategySymbols}, as well as for the combination strategy of FPH + APC.
		Parameters other than $Q_\text{eng}^*$ are not shown for EH, since they are the same as for the base thermonuclear case as they do not enter the internal power balance.
		In the first column are the densities for fast protons, thermal protons, and boron, as well as $Q_\text{eng}^*$ (which is repeated from Fig.~\ref{fig:QVsTauESingle} for ease of reference).
		In the second column are the fast proton energy, and the temperatures of thermal protons, boron, and electrons.
		In the third column are the fusion power from fast protons, fusion power from thermal protons, bremsstrahlung power, and heating power.
	}
	\label{fig:params}
\end{figure*}

\section{Optimal Reactor Parameter Space}

In addition to considering the values of $Q_\text{eng}^*$ and $\tau_E^\dagger$ that result from the optimization, it is also informative to consider what the optimal reactor parameter space looks like.
We focus on the scenario covered in Figs.~\ref{fig:QVsTauESingle}-\ref{fig:TauBreakevenSingle}, to explore the different features of each strategy.
For compactness, the results for all parameters are presented together in the grid in Fig.~\ref{fig:params}, which shows the optimal (for $Q_\text{eng}^*$) densities and temperatures (or energies) of each species as a function of $\tau_E$, as well as the corresponding fast proton fusion power $P_{F,f}$, thermal proton fusion power $P_{F,p}$, bremsstrahlung power $P_B$, and heating power $P_H$.
Results are not shown for EH, since the parameters in this case are the same as for the thermonuclear base case, following the discussion at the beginning of Sec.~\ref{sec:optimization}.

Several trends are apparent from the figure.
First, for the thermonuclear base case, as $\tau_E$ increases, the optimal fraction Boron falls from 60\% to just over 20\%.
It is notable that these Boron fractions are substantially higher than the common mix of 15\% Boron, 85\% protons, which is characteristic of optimizing for ignition.\cite{Putvinski2019,Ochs2022ImprovingFeasibility}
This trend of falling Boron proportion with increasing confinement time persists across all fusion strategies.

Second, for the thermonuclear base case, as $\tau_E$ increases, the temperature of the protons falls and the temperature of boron and electrons increases, as less heating power (which goes into protons) is required to maintain the reaction and less energy is lost. 
This trend holds across all the cases that have only a thermal proton population, however it does \emph{not} hold for the fast proton heating cases.
For these cases, above a certain value of $\tau_E$, the optimal value of $T_p$ actually plunges, and most of the fusion power is provided by the fast particle beam: in fact, the low-$\tau_E$ optimal mix is $\sim 25\%$ fast protons and $\sim 70\%$ Boron, with only $\sim 5\%$ thermal protons.
As $\tau_E$ grows even larger, the optimal thermal proton fraction and temperature both rise, so that eventually in the reactor breakeven scenarios comparable fusion power is produced by thermal and fast protons.

Third, one might have expected DC to be optimized at the lowest electron temperatures, since for DC one desires power to come out as thermal conduction loss (with an electrical conversion efficiency of 85\%) rather than bremsstrahlung (with an electrical conversion efficiency of 64\%), and bremsstrahlung power scales roughly as $T_e^{1/2}$
However, the electron temperatures are in fact the highest for the DC case.
Nevertheless, the bremsstrahlung power is lowest for the DC case as expected, which is accomplished by lowering the fraction of high-$Z$ Boron.

Fourth, one might note that for cases with FPH, the optimal fast proton energy $E_f$ bounces around a bit as a function of $\tau_E$.
It is important here to note that given the broad scan, for the code to run we did not require machine-precision optimality from the solved equilibrium.
Instead, we required only that it be near the optimum, as encoded in the gtol and xtol parameters, which were chosen such that the optimal point should lie within around 10 keV of the discovered solution.
Since the optimum value of the fast ion energy does not change very much, this variation is visible in the final plot.

\section{Summary and Conclusion}

In this paper, we have examined the 0D power balance of a non-igniting pB11 reactor.
For the thermonuclear base case with state-of-the-art thermal conversion, we showed that (before considering the power required to sustain the confinement system) a Lawson product of $n_i \tau_E \approx 1.2 \times 10^{15}$ was required for power plant breakeven, a factor-of-40 improvement over the required product for ignition.
For a typical magnetic confinement fusion density of $n_i \sim 10^{14}$, this implies a 12 second confinement time, rather than a 450 second energy confinement time for an igniting thermonuclear reactor.
Thus, non-igniting pB11 ``wet wood burner''-type reactors, with substantial capture power, are much more easy to envision than the ignited reactors that are typically the target of DT fusion research.

However, 12 seconds is still a formidable energy confinement time, and so we have examined how several strategies, including fast proton heating (FPH), alpha power capture (APC), direct conversion (DC), and efficient heating (EH) might improve reactor feasibility for pB11 fusion. 
Each of these strategies offered significant improvements to the confinement time, as roughly mapped out in Eq.~\ref{eq:linearSensitivity}, with FPH and APC working particularly well in tandem.
A combination of all four strategies reduced the confinement time by almost an order of magnitude, to 1.4 seconds, bringing it in line with the target energy confinement time of mainstream DT fusion experiments such as ITER.
Thus, understanding and harnessing these strategies should play an important role in bringing economical aneutronic fusion closer to reality. 

\section*{Acknowledgments}

This work was supported by ARPA-E Grant DE-AR0001554. 
This work was also supported by the DOE Fusion Energy Sciences Postdoctoral Research Program, administered by the Oak Ridge Institute for Science and Education (ORISE) and managed by Oak Ridge Associated Universities (ORAU) under DOE contract No. DE-SC0014664.




%


\clearpage

\end{document}